\documentclass{aastex631}

\usepackage{amsmath}

\shorttitle{Dissociative Recombination of Rotationally Cold OH$^+$}
\shortauthors{K\'alosi et al.}
\graphicspath{{./}{figures/}}

\begin{document}

\title{Dissociative Recombination of Rotationally Cold OH$^+$ and Its Implications for the Cosmic Ray Ionization Rate in Diffuse Clouds}

\correspondingauthor{\'Abel K\'alosi}
\email{abel.kalosi@outlook.com}

\author[0000-0003-3782-0814]{\'Abel K\'alosi}
\affiliation{Columbia Astrophysics Laboratory, Columbia University, New York, NY 10027, USA}
\affiliation{Max-Planck-Institut f\"ur Kernphysik, Saupfercheckweg 1, D-69117 Heidelberg, Germany}

\author[0009-0002-8746-2728]{Lisa Gamer}
\affiliation{Max-Planck-Institut f\"ur Kernphysik, Saupfercheckweg 1, D-69117 Heidelberg, Germany}

\author[0000-0002-5066-2550]{Manfred Grieser}
\affiliation{Max-Planck-Institut f\"ur Kernphysik, Saupfercheckweg 1, D-69117 Heidelberg, Germany}

\author{Robert von Hahn}
\altaffiliation{Deceased}
\affiliation{Max-Planck-Institut f\"ur Kernphysik, Saupfercheckweg 1, D-69117 Heidelberg, Germany}

\author[0000-0002-3571-5765]{Leonard W. Isberner}
\affiliation{I. Physikalisches Institut, Justus-Liebig-Universit\"at Gie{\ss}en, D-35392 Gie{\ss}en, Germany}
\affiliation{Max-Planck-Institut f\"ur Kernphysik, Saupfercheckweg 1, D-69117 Heidelberg, Germany}

\author[0009-0006-6117-2975]{Julia I. J\"ager}
\affiliation{Max-Planck-Institut f\"ur Kernphysik, Saupfercheckweg 1, D-69117 Heidelberg, Germany}

\author[0000-0003-0511-0738]{Holger Kreckel}
\affiliation{Max-Planck-Institut f\"ur Kernphysik, Saupfercheckweg 1, D-69117 Heidelberg, Germany}

\author[0000-0001-8341-1646]{David A. Neufeld}
\affiliation{Department of Physics \& Astronomy, Johns Hopkins University, Baltimore, MD 21218, USA}

\author[0000-0001-7625-4398]{Daniel Paul}
\affiliation{Columbia Astrophysics Laboratory, Columbia University, New York, NY 10027, USA}
\affiliation{Max-Planck-Institut f\"ur Kernphysik, Saupfercheckweg 1, D-69117 Heidelberg, Germany}

\author[0000-0002-1111-6610]{Daniel W. Savin}
\affiliation{Columbia Astrophysics Laboratory, Columbia University, New York, NY 10027, USA}

\author[0000-0002-6166-7138]{Stefan Schippers}
\affiliation{I. Physikalisches Institut, Justus-Liebig-Universit\"at Gie{\ss}en, D-35392 Gie{\ss}en, Germany}

\author[0000-0002-2808-2760]{Viviane C. Schmidt}
\affiliation{Max-Planck-Institut f\"ur Kernphysik, Saupfercheckweg 1, D-69117 Heidelberg, Germany}

\author[0000-0003-1198-9013]{Andreas Wolf}
\affiliation{Max-Planck-Institut f\"ur Kernphysik, Saupfercheckweg 1, D-69117 Heidelberg, Germany}

\author[0000-0003-0030-9510]{Mark G. Wolfire}
\affiliation{Department of Astronomy, University of Maryland, College Park, MD 20742, USA}

\author[0000-0003-2520-343X]{Old{\v r}ich Novotn\'y}
\affiliation{Max-Planck-Institut f\"ur Kernphysik, Saupfercheckweg 1, D-69117 Heidelberg, Germany}


\begin{abstract}

Observations of OH$^+$ are used to infer the interstellar cosmic ray ionization rate in diffuse atomic clouds, thereby constraining the propagation of cosmic rays through and the shielding by interstellar clouds, as well as the low energy cosmic ray spectrum. In regions where the H$_2$ to H number density ratio is low, dissociative recombination (DR) is the dominant destruction process for OH$^+$ and the DR rate coefficient is important for predicting the OH$^+$ abundance and inferring the cosmic ray ionization rate. We have experimentally studied DR of electronically and vibrationally relaxed OH$^+$ in its lowest rotational levels, using an electron--ion merged-beams setup at the Cryogenic Storage Ring. From these measurements, we have derived a kinetic temperature rate coefficient applicable to diffuse cloud chemical models, i.e., for OH$^+$ in its electronic, vibrational, and rotational ground level. At typical diffuse cloud temperatures, our kinetic temperature rate coefficient is a factor of $\sim 5$ times larger than the previous experimentally derived value and a factor of $\sim 33$ times larger than the value calculated by theory. Our combined experimental and modelling results point to a significant increase for the cosmic ray ionization rate inferred from observations of OH$^+$ and H$_2$O$^+$, corresponding to a geometric mean of $(6.6 \pm 1.0) \times 10^{-16}\,\mathrm{s}^{-1}$, which is more than a factor of two larger than the previously inferred values of the cosmic ray ionization rate in diffuse atomic clouds. Combined with observations of diffuse and dense molecular clouds, these findings indicate a greater degree of cosmic ray shielding in interstellar clouds than has been previously inferred.

\end{abstract}

\section{Introduction} \label{sec:intro}

Cosmic rays are  an important component of our Galaxy. In the local Galaxy their total energy density ($\sim 1\, \mathrm{eV}\, \mathrm{cm}^{-3}$) is comparable to that of star light, the magnetic field, and the cosmic microwave background \citep{grenier_nine_2015}. Cosmic rays are the dominant ionization source for atomic and molecular hydrogen in the cold neutral medium (CNM). Through this ionization they play a central role in initiating astrochemistry in the diffuse atomic and molecular  and dense molecular  phases of the interstellar medium \citep[ISM;][]{gerin_hydrides_2016, indriolo_cosmic_2013, hollenbach_cheminter_2012}. In addition, the ionization provides coupling with the magnetic field, and in molecular cloud cores, slows down core collapse and inhibits both star and disk formation \citep{mckee_star_2007, padovani_cosmic_2018}. Thus, the cosmic ray ionization rate (CRIR) is an important parameter for both chemical and dynamical models in these environments.

Observations have indicated that cosmic ray shielding decreases the ionization rate with increasing column density into a cloud. The rates estimated from low column density diffuse clouds \citep[e.g.,][]{neufeld_cosmic_2017} are higher than those found in dense cloud interiors \citep{caselli_ionization_1998, vandertak_limits_2000} by factors of $\sim 10$. Here we focus on diffuse clouds in order to obtain an ionization  rate at cloud surfaces that is largely unshielded. The ionization rate at the cloud surface is one of the key parameters needed to model the astrochemistry in the ISM. In addition, comparing the surface rate to those estimated at larger columns acts as a guide to future researchers in understanding the details of cosmic ray shielding.

The CRIR in diffuse clouds is commonly inferred through observations of molecular cations such as H$_{3}^{+}$, OH$^+$, and ArH$^+$ \citep{indriolo_investigating_2012, gerin_hydrides_2016, neufeld_cosmic_2017}. These are associated with specific local fractions of molecular hydrogen $f_{\mathrm{H}_{2}} \equiv 2n(\mathrm{H}_{2})/n_{\mathrm{H}}$, where $n(\mathrm{H}_2)$ is the number density of molecular hydrogen and $n_{\mathrm{H}}$ is the total hydrogen nuclei number density. H$_{3}^{+}$ is found both in dense and diffuse molecular clouds \citep[e.g.,][]{mccall_enhanced_2003, brittain_inter_2004, indriolo_diffuse_2007}. OH$^+$ probes the mostly atomic layer of diffuse clouds with $f_{\mathrm{H}_{2}} \sim 0.1$ \cite[e.g.,][]{indriolo_chemical_2012}. ArH$^+$ is found in almost purely atomic diffuse clouds with $f_{\mathrm{H}_{2}} < 0.01$ \citep[e.g.,][]{schilke_argonium_2014}. Here, we focus on OH$^+$ and the astrochemistry relevant to inferring the CRIR from the observed abundances.

The relative simplicity of OH$^+$ chemistry in diffuse clouds makes it a powerful probe of the CRIR \citep{hollenbach_cheminter_2012}. Cosmic ray ionization of H forms H$^+$, which can then undergo nearly resonant charge transfer with O to form O$^+$. This can then be followed by an exoergic hydrogen abstraction reaction to form OH$^+$ via
\begin{equation}
    \mathrm{O}^+ + \mathrm{H}_2 \rightarrow \mathrm{OH}^+ + \mathrm{H}.
\end{equation}
The direct link of OH$^+$ to the cosmic ray ionization of H makes its abundance roughly proportional to the CRIR over a wide range of parameter space.\footnote{The formation pathway via H$_2^+$ is not considered since OH$^+$ in diffuse clouds is expected to peak at low $f_{\mathrm{H}_{2}}$ \citep{hollenbach_cheminter_2012} where this pathway is not important.} OH$^+$ can be destroyed by hydrogen abstraction reaction, forming H$_2$O$^+$ via
\begin{equation}\label{eq:ohabs}
    \mathrm{OH}^+ + \mathrm{H}_2 \rightarrow \mathrm{H}_2\mathrm{O}^+ + \mathrm{H}
\end{equation}
and also by dissociative recombination (DR) with free electrons via
\begin{equation}\label{eq:dr}
    \mathrm{OH}^+ + \mathrm{e}^- \rightarrow \mathrm{O} + \mathrm{H}.
\end{equation}
Recent laboratory studies have measured the hydrogen abstraction chain for diffuse cloud temperatures with an accuracy of $20\%$ \citep{tran_formation_2018, kovalenko_formation_2018, kumar_rates_2018}. \cite{neufeld_cosmic_2017} have identified the OH$^+$ DR rate coefficient (i.e., kinetics) as one of the key uncertainties impacting the inferred CRIR using OH$^+$ chemistry in diffuse clouds.

The reliability of the OH$^+$ DR rate coefficient currently listed in astrochemical databases is a significant issue of concern. In diffuse clouds, OH$^+$ is predicted to be in its lowest electronic, vibrational, and rotational level, neglecting its hyperfine structure. The KIDA database \citep{wakelam_kida_2012} has adopted the theoretical work of \cite{guberman_diss_1995} for ground rotational level OH$^+$. A later theoretical study by \cite{stroe_electron_2018} expanded the calculations of \cite{guberman_diss_1995} by including core-excited Rydberg states into the existing theoretical framework and reproduced the previous results to within $20\%$. Quantum mechanical calculations are extremely challenging due to the many-body nature of the problem and the infinite number of intermediate states involved in the DR process. For example, the recent unified theoretical treatment of DR by \cite{forer_unified_2023} for CH$^+$ reproduce the experimental cross section results of \cite{paul_dr_2022} only to within a factor of 2 to 5. Further theoretical improvements are required to reach experimental accuracy. As a result, laboratory measurements are still expected to be the most reliable means to generate DR kinetics data. The UMIST database \citep{mcelroy_umist_2012} and most astrochemical modelers follow this recommendation and have adopted the experimental results of \cite{mitchell_diss_1990}, which are a factor of $6$ larger than the calculations of \cite{guberman_diss_1995}. However, the rate coefficient of \cite{mitchell_diss_1990} was derived from single-pass merged-beams measurements for which the OH$^+$ ions were electronically, vibrationally, and rotationally excited. \cite{amitay_oh_1996} studied the DR of electronically and vibrationally relaxed OH$^+$ at the room-temperature Test Storage Ring (TSR). Their measurement, though, was only on a relative scale.  Moreover, the rotational level population in the experiment was most likely close to room temperature. Recent experimental work for DR of HeH$^+$ and CH$^+$ has shown that for CNM temperatures the DR rate coefficient for internally excited ions can be over an order of magnitude larger or smaller than that for fully relaxed ions \citep{novotny_quantum_2019, paul_dr_2022}. One aim of our work here is to generate DR kinetics data for internally cold OH$^+$ to an accuracy of $\sim 20\%$, so that any remaining discrepancies between diffuse cloud chemical models and observations cannot be attributed to uncertainties in the underlying chemistry but rather begin to tell us about the astrophysics of diffuse clouds.

OH$^+$ in diffuse clouds was first detected via absorption by rotational lines in the far infrared \citep[IR;][]{wyrowski_firstinterstellar_2010, gerin_interstellarOH_2010, neufeld_herschelhifi_2010} and by electronic transitions in the near ultraviolet \citep[UV;][]{krelowski_hydroxilUV_2010}. Further near-UV observations found CRIR estimates similar to those derived from far-IR lines \citep{porras_diffuse_2013, zhao_translucent_2015}. A subsequent IR survey by \cite{indriolo_survey_2015} detected OH$^+$, H$_2$O$^+$, and H$_3$O$^+$, providing one of the largest samples to date and demonstrating the usefulness of oxygen-bearing ions for probing the CRIR of hydrogen. The ratio of the OH$^+$ to H$_2$O$^+$ column densities was used to infer $f_{\mathrm{H}_{2}}$, an important parameter in determining the CRIR. \cite{bacalla_edibles_2019} extended and re-evaluated the previous near-UV OH$^+$ observations, implementing an update of the electronic-band line oscillator strengths and an updated chemical model. Their results suggest a somewhat higher CRIR compared to the far-IR studies. However, the near-UV studies lack corresponding H$_2$O$^+$ observations, requiring those studies to assume a value for $f_{\mathrm{H}_{2}}$ in order to model the OH$^+$ chemistry. In addition, both the far-IR and near-UV studies used a single-zone chemical model to interpret the OH$^+$ observations, leading to the approximation that OH$^+$ and H$_2$O$^+$ exist in the cloud at the same location and set of physical conditions.

Some researchers have developed one-dimensional (1D) models as a function of visual extinction in diffuse clouds, so as to be able to more reliably interpret the simultaneous observation of multiple molecules within a diffuse cloud \citep{hollenbach_cheminter_2012, neufeld_cheminterarg_2016, neufeld_cosmic_2017}. These models predict that OH$^+$, H$_2$O$^+$, and H$_3$O$^+$ peak in abundance at different depths into the cloud, contrary to the assumptions of single-zone models. However, a full comparison of the observed and modelled column densities requires reliable DR data for all involved species, most importantly for OH$^+$ and H$_2$O$^+$.

Here, we report our measurements for the DR rate coefficient of OH$^+$ in its ground electronic, vibrational, and rotational level. Our results are applicable to the chemistry of OH$^+$ in the ISM over a broad range of kinetic temperatures. In addition, we assess the impact of our data on models of diffuse cloud chemistry and their subsequent implication for deriving the CRIR from astronomical observations. As for DR of H$_2$O$^+$, room-temperature storage ring experiments have measured DR of electronically and vibrationally relaxed H$_2$O$^+$ ions, but the rotational level population was approximately room temperature  \citep{jensen_h2o_1999, rosen_h2o_2000}. We are unaware of any DR measurements for rotationally relaxed H$_2$O$^+$ and consider it a candidate for a future DR study.

\section{Experiment} \label{sec:exp}

DR measurements were performed on internally cold OH$^+$ using the Cryogenic Storage Ring \citep[CSR;][]{vonhahn_csr_2016} facility at the Max Planck Institute for Nuclear Physics in Heidelberg, Germany. The methodology for performing DR measurements at CSR has been described in detail elsewhere \citep{novotny_quantum_2019, paul_dr_2022}. Here we provide only a brief overview, with an emphasis on those aspects that are most specific to the present results. Additional details can be found in the Appendices.

OH$^+$ ions were generated in a gas-discharge source, producing internally excited ions that were injected into CSR  (see Appendix~\ref{app:mbexp}). The stored ions rapidly relaxed to their ground electronic and vibrational states. Rotationally, the ions relaxed via radiative interactions toward thermal equilibrium with the cryogenic black-body temperatures of CSR \citep{oconnor_photo_2016, meyer_radiative_2017, kalosi_inelastic_2022}. The resulting rotational distribution was much colder than that for previous OH$^+$ studies using room-temperature storage rings \citep{amitay_oh_1996, stromholm_imaging_1997, hechtfischer_nearthreshold_2019}. Additional rotational cooling was provided through electron--ion rotational-level-changing collisions using the electron beam described below. Using the collisional-radiative model of \cite{kalosi_inelastic_2022} for the populations of the rotational levels, labelled by quantum number $N$, we estimate that $\approx 80\%$ of the ions were in the $N=0$ ground level and $\approx 20\%$ in the $N=1$ level (see Appendix~\ref{app:rotmodel}).

DR measurements were performed using an electron--ion merged-beams configuration that is discussed in more detail in Appendix~\ref{app:mbexp}. We merged a magnetically guided electron beam onto the stored ions in one of the straight sections of CSR. The relative collision energy between the merged beams was controlled by tuning the nominal laboratory-frame electron-beam energy in the interaction region, $E_\mathrm{e}$. DR events resulted in neutral fragments, see Reaction~(\ref{eq:dr}), at nearly the same laboratory-frame velocity as the initial ion. The resulting neutral reaction products were collected by a particle-counting detector downstream from the interaction region. The recorded count rate is proportional to the merged-beams rate coefficient
\begin{equation}
    \alpha^\mathrm{mb} = \langle \sigma v \rangle,
\end{equation}
where $\sigma$ is the energy dependent DR cross section, $v$ is the electron--ion collision velocity, and the angled brackets designate the average over the velocity distribution in the electron--ion overlap region. The velocity distribution is determined by several experimental factors: the perpendicular and parallel  temperature components $T_\perp$ and $T_\parallel$, respectively, relative to the bulk electron-velocity vector; the merging geometry; and the variable laboratory-frame energy along the interaction region. The resulting velocity distribution is modelled following the approach of \cite{novotny_drhcl_2013}, as extended by \cite{kalosi_inelastic_2022}.  This velocity distribution differs significantly from the Maxwell--Boltzmann (i.e., kinetic-temperature-dependent) distribution used in astrochemical models. Extraction of the underlying cross section and subsequent conversion of the data is required before it can be applied in kinetic models \citep[e.g.,][]{paul_dr_2022}. The velocity distribution can also be translated into a collision-energy distribution as a function of the detuning energy $E_\mathrm{d}$, defined as the nominal center-of-mass collision energy, 
\begin{equation}\label{eq:det}   
    E_\mathrm{d} = \left(\sqrt{E_\mathrm{e}} - \sqrt{E_0} \right)^2,
\end{equation}
where $E_0$ is the laboratory-frame electron-beam energy at matched electron--ion velocities, taking into account that the reduced mass of the collision system is essentially equal to that of the electron. We use the full width at half maximum (FWHM) of the collision-energy distribution as a measure of the energy resolution in the experiment. The evaluation of the experimental data and the absolute scaling of the merged-beams results are discussed in Appendix~\ref{app:absscale}.

Measurements were performed using a nearly pure beam of OH$^+$. To achieve this, we operated CSR in the recently developed isochronous mode, i.e., as a time-of-flight mass spectrometer with a mass resolution of $\sim 10^{-5}$~u \citep{grieser_iso_2022}. A particular concern was potential contamination with NH$_3^+$, which differs in mass from OH$^+$ by $\approx 2\times10^{-2}$~u. For the present results, we find that $< 0.1\%$ of the ion beam was due to NH$_3^+$, using the methods developed by \cite{grieser_iso_2022}.

\section{Experimental Results} \label{sec:res}

\begin{figure*}[ht!]
\plotone{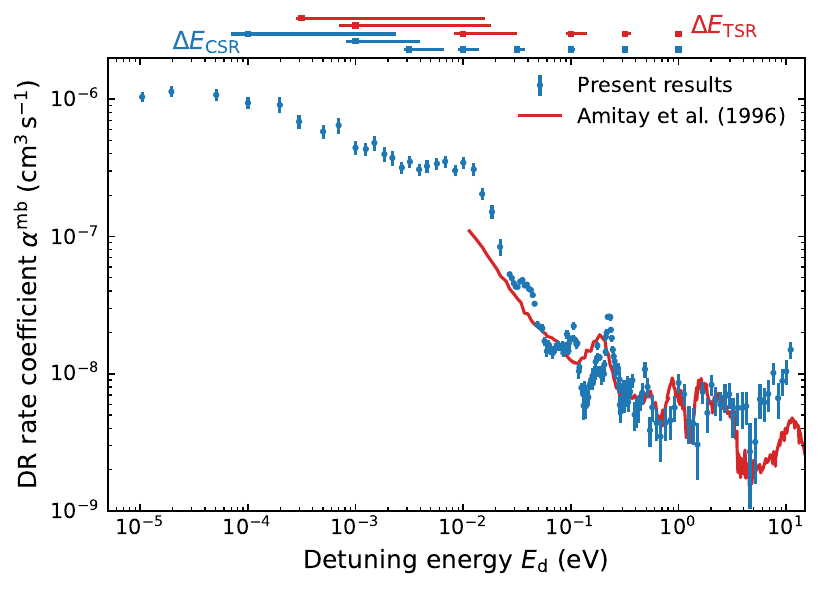}
\caption{Experimental merged-beams DR rate coefficient $\alpha^\mathrm{mb}(E_\mathrm{d})$ for stored OH$^+$ ions. The present results are plotted as blue symbols with error bars representing one-sigma statistical uncertainties. The absolute scaling of the results has a systematic accuracy of $17\%$ at all energies. For comparison, the relative results of \cite{amitay_oh_1996} from the room-temperature TSR study are added as the full red line, after applying a common scaling factor at all energies to their values to best match the present data in the $1$ to $3$~eV region. For $E_\mathrm{d} > 5$~eV, our results may contain an additional contribution from DE, as is discussed in the text. At the top of the figure, we have plotted the energy-dependent energy resolution $\Delta E$ for both experiments. For the CSR results, the energy resolution, shown in blue, is calculated as the FWHM of the simulated collision-energy distribution. For the TSR results, we calculated the energy resolution, shown in red, using the FWHM from the experimental parameters of \cite{amitay_oh_1996}. The square symbols show selected values of $E_\mathrm{d}$ and the horizontal lines the FWHM region of the corresponding collision-energy distributions. \label{fig:mbrc}}
\end{figure*}

The measured merged-beams rate coefficient $\alpha^\mathrm{mb}(E_\mathrm{d})$ for DR of OH$^+$ is plotted in Figure~\ref{fig:mbrc}, showing a rich structure of features. We can compare our results with those of \cite{amitay_oh_1996}, which were measured at the room-temperature TSR. At the top of Figure~\ref{fig:mbrc}, we have plotted the energy-dependent energy resolution $\Delta E$ for the CSR and TSR results. At the lowest energies, the energy resolution for the CSR results is almost an order of magnitude finer than for the TSR data. In order to qualitatively compare our work with the TSR results, we have scaled their relative values with a common factor at all energies to best match our data in the $1$ to $3$~eV region. Between $\sim 0.1$ and 1~eV, we find reasonable agreement in the shape of the CSR and TSR DR results, taking into account the differences in the energy resolution of the two data sets. Below $\sim 0.1$~eV, the CSR results show structure that is not seen in the TSR data. We attribute these differences to the improved energy resolution of the CSR data and to the well defined and lower internal excitation of the ions, as discussed below. Above $\approx 5$~eV, approximately corresponding to the dissociation energy of OH$^+$ \citep{hechtfischer_nearthreshold_2019}, the CSR and TSR results diverge. We attribute this to the opening of the dissociative excitation (DE) channels
\begin{equation}\label{eq:de}
    \mathrm{OH}^+ + \mathrm{e}^- \rightarrow \begin{cases}
    \mathrm{O}^+ + \mathrm{H} + \mathrm{e}^- \\
    \mathrm{O} + \mathrm{H}^+ + \mathrm{e}^-.
    \end{cases}
\end{equation}
The CSR DR detection system is currently not capable of distinguishing between DR and DE, while the TSR experiments used a mass-sensitive detection method that enabled \cite{amitay_oh_1996} to select for DR and discriminate against DE events.

The astrochemical relevance of our DR measurements is due to three advances over the TSR results, which we discuss in no particular order. First, the order of magnitude finer energy spread in the present experiment enables us to access collision energies relevant to diffuse cloud conditions. This improved energy resolution is most clearly demonstrated by the resonant feature observed below $20$~meV. Second, we have measured $\alpha^\mathrm{mb}(E_\mathrm{d})$ on an absolute scale, to be contrasted with the relative values from TSR, and have converted it into a kinetic-temperature-dependent rate coefficient, $\alpha^\mathrm{k}(T_\mathrm{k})$, for use in chemical models (see Appendix~\ref{app:ratecoeff}). Here, $T_{\mathrm k}$ is the kinetic temperature of the gas, which characterizes the velocity distribution for all particles. Lastly, is our ability to generate internally cold OH$^+$ ions, similar to the conditions expected in diffuse clouds. Our model for the level populations predicts an $N=0$ relative population during the measurement of $(80\pm5) \%$ and $N=1$ of $(19\pm4) \%$, where here and throughout all uncertainties are given at a one-sigma confidence level (see Appendix~\ref{app:rotmodel}). This contrasts with the room-temperature TSR results, where these two levels are predicted to have relative populations of $8 \%$ and $20 \%$, respectively. In diffuse clouds, the OH$^+$ is predicted to be nearly $100 \%$ in the $N=0$ level. The critical density for collisions of OH$^+$ with H and e$^-$ for the $N=0 \rightarrow N=1$ transition are calculated to be on the order of $\approx10^{7}$~cm$^{-3}$ and $\approx10^{4}$~cm$^{-3}$, respectively, which are orders of magnitude larger than the estimated H and e$^-$ densities in diffuse clouds. Our storage-time dependent DR measurements, similar to the studies for CH$^+$ of \cite{paul_dr_2022}, enable us to probe the changing contributions of the OH$^+$ $N=0$ and $N=1$ levels  (see Appendix~\ref{app:vststorm}). We find that the differences are most noticeable for $E_\mathrm{d} < 30$~meV but are negligible when converting $\alpha^\mathrm{mb}$ to $\alpha^\mathrm{k}$.

\begin{figure*}[ht!]
\plotone{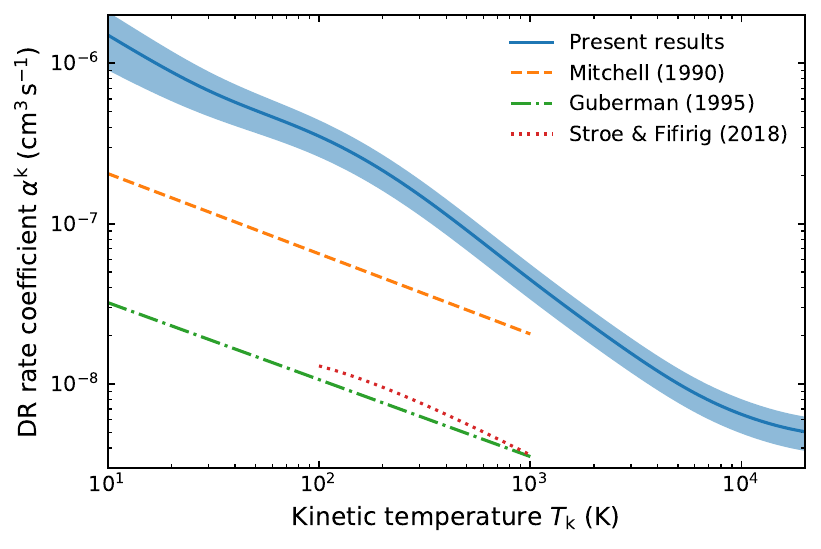}
\caption{Comparison of the kinetic-temperature-dependent DR rate coefficient $\alpha^\mathrm{k}$ from our present experiment to previously published works. The shaded area around the present results corresponds to the total systematic uncertainty of the measurement, mainly due to the absolute scaling of $\alpha^\mathrm{mb}$ and the uncertainty of $T_\perp$. The particular contributions of the uncertainties have been discussed by \cite{paul_dr_2022}. The single-pass merged-beams results of \cite{mitchell_diss_1990} have been incorporated into the UMIST database \citep{mcelroy_umist_2012}. The theoretical results of \cite{guberman_diss_1995} have been incorporated into the KIDA database \citep{wakelam_kida_2012}.\label{fig:tkin}}
\end{figure*}

We have generated $\alpha^\mathrm{k}(T_\mathrm{k})$ from our $\alpha^\mathrm{mb}(E_\mathrm{d})$ results, following the DR cross-section-extraction method of \cite{novotny_drhcl_2013} with further improvements from \cite{paul_dr_2022}. We have determined $\alpha^\mathrm{k}$ from $T_\mathrm{k} = 10$ to $20,\!000$~K, as plotted in Figure~\ref{fig:tkin}. Simple fitting formulae for the rate coefficient are given in Appendix~\ref{app:ratecoeff}. Also shown in the figure are the literature values from the single-pass merged-beams experiment of \cite{mitchell_diss_1990} and the theoretical calculation of \cite{guberman_diss_1995}. These DR rate coefficients have been adopted by the astrochemistry databases UMIST \citep{mcelroy_umist_2012} and KIDA \citep{wakelam_kida_2012}, respectively. Figure~\ref{fig:tkin} also shows the results of \cite{stroe_electron_2018}, which agree with that of \cite{guberman_diss_1995} to within $20\%$. At typical diffuse cloud temperatures of $40-130$~K \citep{shull_faruv_2021}, our kinetic temperature rate coefficient is a factor of $\sim 5$ times larger than the experimentally derived value of \cite{mitchell_diss_1990} and a factor of $\sim 33$ times larger than the theoretical results of \cite{guberman_diss_1995}.

We also note that the kinetic temperature rate coefficients of \cite{mitchell_diss_1990} and \cite{guberman_diss_1995} exhibit a nearly $T_\mathrm{k}^{-1/2}$ behavior. However, our experimentally derived value has a slope that varies significantly with temperature between $T_\mathrm{k}^{-1/2}$ and $T_\mathrm{k}^{-1}$. A variable slope with temperature has also been seen by \cite{novotny_quantum_2019} for HeH$^+$ and by \cite{paul_dr_2022} for CH$^+$. Together with those works, our findings demonstrate that the DR kinetic temperature rate coefficient can exhibit a temperature dependence that differs significantly from the theoretically derived $T_\mathrm{k}^{-1/2}$ behavior for direct DR of diatomic ions \citep{guberman_diss_1995}.

\section{Astrophysical Implications} \label{sec:astro}

We have investigated the astrophysical impact of our DR rate coefficient $\alpha^\mathrm{k}$ using the single-zone model of \cite{bacalla_edibles_2019} and an updated version of the one-dimensional (1D) model of \cite{neufeld_cosmic_2017}. These translate the observed OH$^+$ column densities $N(\mathrm{OH}^+)$ into an inferred CRIR. Both studies have adopted similar reaction networks and identified the same dominant destruction pathways for OH$^+$ in diffuse clouds, i.e., Reactions~(\ref{eq:ohabs}) and (\ref{eq:dr}).

\begin{figure*}[ht!]
\plotone{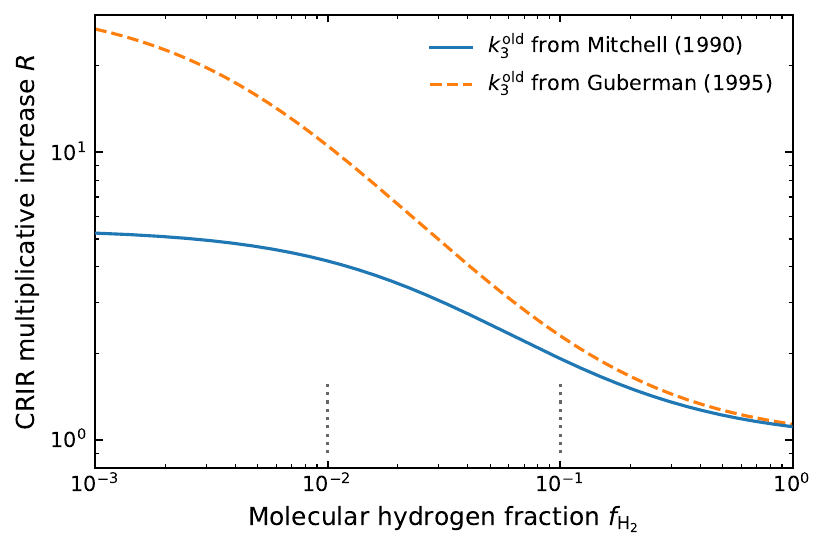}
\caption{Increase of the inferred CRIR from OH$^+$ observations in diffuse clouds using our updated DR rate coefficient for single-zone models. The multiplicative increase $R$ is calculated from the ratio of the destruction rates (see Eq.~\ref{eq:ratio}) for the present ($\alpha^\mathrm{k}$) and the previous ($k_3^\mathrm{old}$) DR rate coefficients, adopting the single-zone model of \cite{bacalla_edibles_2019}, which use the DR data of \cite{mitchell_diss_1990} for $k_3^\mathrm{old}$. The result is plotted as the blue full line. The vertical dotted lines enclose the relevant range of $f_{\mathrm{H}_{2}}$ values inferred by \cite{indriolo_survey_2015}. The dashed orange line shows the CRIR increase relative to using the DR rate coefficient of \cite{guberman_diss_1995} for $k_3^\mathrm{old}$.
\label{fig:dest}}
\end{figure*}

To demonstrate the impact of the present DR rate coefficient on the inferred CRIR values, we have calculated the destruction rate of OH$^+$ due to DR and hydrogen abstraction as a function of $f_{\mathrm{H}_{2}}$, similar to the methods of \cite{indriolo_survey_2015} and \cite{bacalla_edibles_2019}. The resulting inferred CRIR of atomic hydrogen $\zeta_\mathrm{H}$ is proportional to the sum of destruction rates
\begin{equation} \label{eq:crir}
    \zeta_\mathrm{H} \propto \left( x_\mathrm{e}k_3 + \frac{f_{\mathrm{H}_2}}{2}k_2 \right),
\end{equation}
where $x_\mathrm{e} = n(\mathrm{e})/n_\mathrm{H}$ is the electron fraction, and $k_2$ and  $k_3$ are the rate coefficients for Reactions~(\ref{eq:ohabs}) and (\ref{eq:dr}), respectively. For further details on Equation~(\ref{eq:crir}), see \cite{indriolo_survey_2015}. The ratio of the destruction rates
\begin{equation} \label{eq:ratio}
    R = \frac{\left( x_\mathrm{e}\alpha^\mathrm{k} + \frac{f_{\mathrm{H}_2}}{2}k_2 \right)}{\left( x_\mathrm{e}k_3^\mathrm{old} + \frac{f_{\mathrm{H}_2}}{2}k_2 \right)},
\end{equation}
calculated with the present ($k_3 = \alpha^\mathrm{k}$, given in Appendix~\ref{app:ratecoeff}) and the previous ($k_3 = k_3^\mathrm{old}$) DR rate coefficients, multiplicatively increases the CRIR values inferred by \cite{bacalla_edibles_2019}.\footnote{We do not apply the same treatment to the analysis of \cite{indriolo_survey_2015}, as they did not include a detailed treatment of OH$^+$ formation but instead assumed that a fixed fraction of CR ionizations led to the formation of OH$^+$. Here, we treat the formation of OH$^+$ using the updated model of \cite{neufeld_cosmic_2017}.} The resulting factor is shown as a function of $f_{\mathrm{H}_{2}}$ in Figure~\ref{fig:dest}, using the model parameters of \cite{bacalla_edibles_2019}. They adopted the DR rate coefficient of \cite{mitchell_diss_1990}, which we use in Eq.~\ref{eq:ratio} for $k_3^\mathrm{old}$. For the value of $k_2$, we use $1\times10^{-9}$ from \cite{jones_rate_1981}, which has been incorporated into the UMIST database and adopted by \cite{indriolo_survey_2015} and \cite{bacalla_edibles_2019}. This value does not significantly differ from the results of \cite{tran_formation_2018} and \cite{kumar_rates_2018}, and we use it to stay consistent with the aforementioned astronomical studies. We also show the multiplicative increase using the data of \cite{guberman_diss_1995} for $k_3^\mathrm{old}$.

\begin{figure*}[ht!]
\plotone{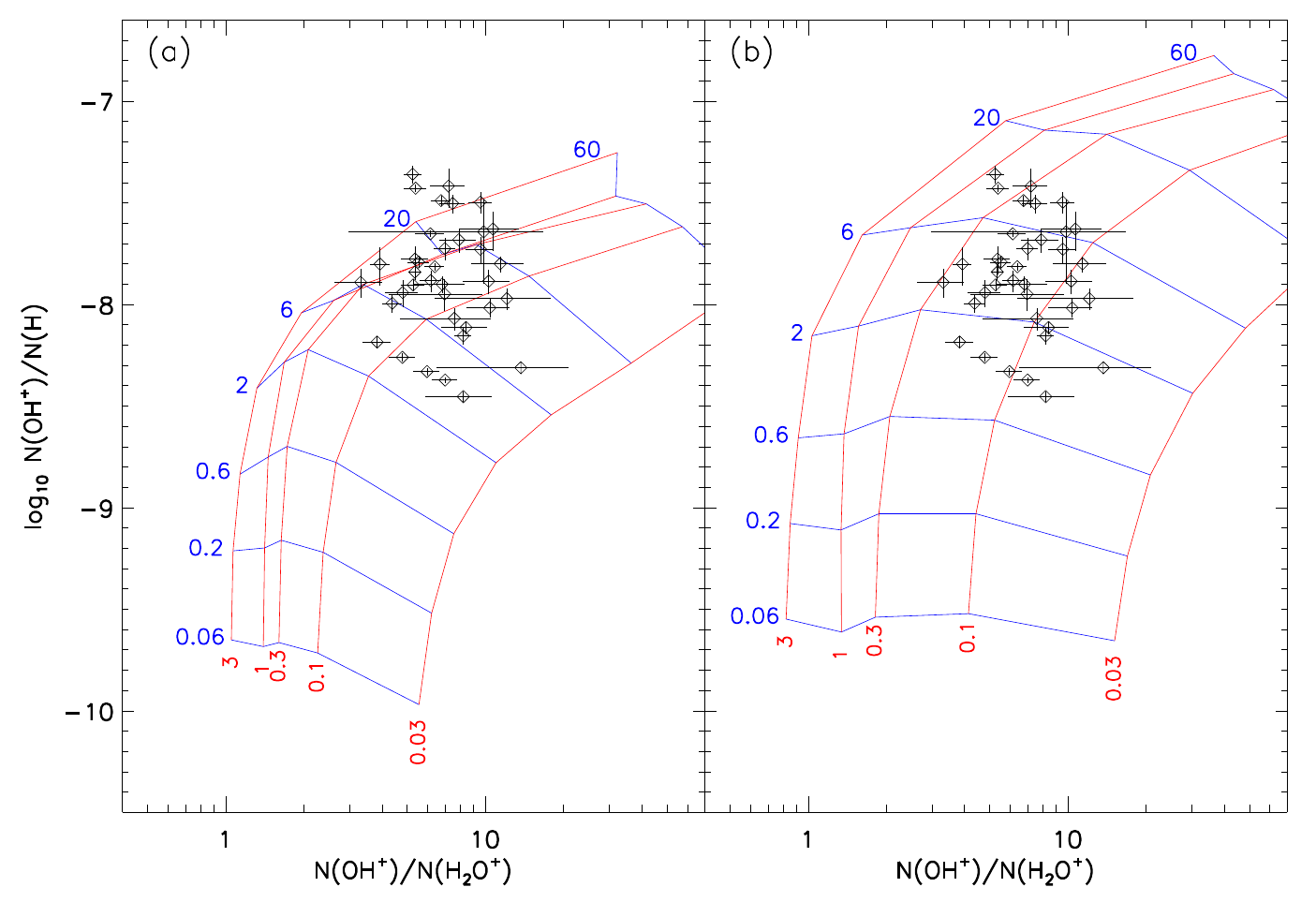}
\caption{Comparison of diffuse cloud model results with the observations of \cite{indriolo_survey_2015} in the plane of the column density ratios $N({\mathrm{OH}}^+)/N({\mathrm{H}_2\mathrm{O}^+})$ and $N({\mathrm{OH}}^+)/N({\mathrm{H}})$. The observations of \cite{indriolo_survey_2015} are plotted as diamond symbols with error bars. The model results are plotted as the blue and red curves. The blue curves are loci of constant $\zeta_p({\mathrm{H}})/n_{50}$ in units of $10^{-16}$~s$^{-1}$ labelled by their corresponding values. The red curves are loci of constant $A_\mathrm{V}({\mathrm{tot}})$ in magnitudes labelled by their corresponding values. (a) Predictions of the present model. (b) Predictions of \cite{neufeld_cosmic_2017}.
\label{fig:crir}}
\end{figure*}

We find a factor of $1.6$ increase for the inferred CRIR values of \cite{bacalla_edibles_2019}, who assumed a single value of $f_{\mathrm{H}_{2}} \approx 0.17$. However, their $f_{\mathrm{H}_{2}}$ value is likely to be an overestimate when contrasted with the IR observations of \cite{indriolo_survey_2015}, who used the ratio of the OH$^+$ to H$_2$O$^+$ column densities to infer $f_{\mathrm{H}_{2}}$ and found $f_{\mathrm{H}_{2}} < 0.1$ for the majority of observations. The above quoted values of $f_{\mathrm{H}_{2}}$ represent the average for those parts of a cloud where OH$^+$ is most abundant. We do not use values for $f_{\mathrm{H}_{2}}$ determined from observed H and H$_2$ column densities \citep[e.g.,][]{winkel_hydrogen_2017} because those measurements are dominated by regions where the OH$^+$ abundance is insignificant. The impact of our present results on the interpretation of the \cite{bacalla_edibles_2019} observations is even greater when values of $f_{\mathrm{H}_{2}} < 0.17$ are considered. Complementary observations of H$_2$O$^+$ for the sightlines in their work do not yet exist but would enable one to better determine the appropriate value of $f_{\mathrm{H}_{2}}$ and thereby make a more reliable comparison between the near-UV and far-IR observations. We note that the $f_{\mathrm{H}_{2}}$ values inferred by \cite{indriolo_survey_2015} directly depend on the DR rate coefficient of H$_2$O$^+$. The uncertainty in this rate coefficient represents a multiplicative scaling factor with larger values leading to an increase in the inferred $f_{\mathrm{H}_{2}}$ and smaller values leading to a decrease.

Moving on to 1D diffuse cloud models, we find that there is a significant impact from our DR rate coefficient on the CRIR inferred for diffuse atomic clouds. We have redetermined the CRIR implied by the OH$^+$ and H$_2$O$^+$ observations presented by \cite{indriolo_survey_2015}. Our modeling results are compared directly to the reported column densities for OH$^+$, H$_2$O$^+$, and atomic H ($N({\mathrm{OH}}^+)$, $N({\mathrm{H}_2\mathrm{O}}^+)$, and $N({\mathrm{H}})$, respectively), and are independent of the single-zone model and from the $f_{\mathrm{H}_{2}}$ values of \cite{indriolo_survey_2015}. To facilitate a comparison with observations of H$_3^+$, we report below the primary CRIR per H atom, $\zeta_p({\mathrm{H}})$. The inferred total CRIR $\zeta_\mathrm{H}$ additionally includes secondary ionizations by energetic electrons and depends on the ionization fraction and $f_{\mathrm{H}_2}$ \citep{dalgarno_electron_1999}. Here, the new OH$^+$ DR rate coefficient obtained in the present study was incorporated into the \cite{neufeld_cosmic_2017} model, along with several other updates: these include the use of (1) new estimates of the rates for several photoprocesses and their depth dependence \citep{heays_photo_2017}; (2) a revised treatment of the heating rate in diffuse clouds\footnote{In \cite{hollenbach_cheminter_2012} an additional heating rate was added to our code to account for photodetachment of electrons from PAH$^-$ by optical photons. We have recently realized that this heating process was already accounted for in the \cite{bakes_photo_1994} formula that we adopt for grain photoelectric heating. For a standard diffuse cloud with $\chi_\mathrm{UV} = 1$, $n_\mathrm{H} = 50$~cm$^{-3}$, and $A_\mathrm{V}({\mathrm{tot}}) = 0.5$ we find that the gas temperature decreases by $13\%$ when removing this extra heating rate.}; and (3) new measurements of the rate coefficients \citep{kovalenko_formation_2018, tran_formation_2018} for the reactions of O$^+$, OH$^+$ and H$_2$O$^+$ with H$_2$. Interpreted with this updated version of the \cite{neufeld_cosmic_2017} grid of 1D models for diffuse clouds, the OH$^+$ and H$_2$O$^+$ column densities measured by \cite{indriolo_survey_2015} imply an average of the logarithmic values $\mathrm{log}_{10}\,\zeta_p({\mathrm{H}})/[1\,\mathrm{s}^{-1}] = (-15.18 \pm 0.06)$, where $\zeta_p({\mathrm{H}})$ is the primary CRIR per H atom. This average value corresponds to the geometric mean $\zeta_p({\mathrm{H}}) = (6.6 \pm 1.0) \times 10^{-16}\,\mathrm{s}^{-1}$. The stated uncertainties are purely statistical in nature and represent the standard error on the mean. Our new estimate of the CRIR is a factor of 3 times as large as that obtained by \cite{neufeld_cosmic_2017}. A factor $2$ increase is directly attributable to the larger DR rate coefficient obtained in the present study, in agreement with the expectations indicated in Figure~\ref{fig:dest}, with the remaining factor of $1.5$ being the result of all the other updates to the model.

The changes are represented graphically in Figure~\ref{fig:crir}, where example model results are shown in the plane of the two observable quantities: the column density ratios $N({\mathrm{OH}}^+)/N({\mathrm{H}_2\mathrm{O}^+})$ and $N({\mathrm{OH}}^+)/N({\mathrm{H}})$. The diamond symbols with error bars show the values measured by \cite{indriolo_survey_2015}, while the red and blue curves show the model predictions as a function of $\zeta_p({\mathrm{H}})/n_{50}$ -- where $n_{50} = n_\mathrm{H}/[50\,\mathrm{cm}^{-3}]$ -- and the total visual extinction through the cloud, $A_\mathrm{V}({\mathrm{tot}})$. Here, the blue curves are loci of constant $\zeta_p({\mathrm{H}})/n_{50}$ in units of $10^{-16}$~s$^{-1}$; and the red curves are loci of constant $A_\mathrm{V}({\mathrm{tot}})$ in magnitudes. Figure~\ref{fig:crir}(a) shows the present model predictions, while Figure~\ref{fig:crir}(b) shows the \cite{neufeld_cosmic_2017} predictions (as shown previously in \cite{neufeld_cosmic_2017}, Figure~7). The downward displacement of the blue curves in Figure~\ref{fig:crir}(a) with respect to Figure~\ref{fig:crir}(b) is the result of our modifications to the \cite{neufeld_cosmic_2017} model and is the explanation for why our revised estimate of the CRIR is three times as large.

We note that our results depend on the $\mathrm{H}^+ + \mathrm{O}$ charge exchange rate coefficient and there is some disagreement in the literature on the predicted value. The data used here are from \cite{stancil_charge_1999} which are a factor of $\sim 6$ larger than those of \cite{spirko_charge_2003}. Here, using the new OH$^+$ DR rate coefficient and the \cite{spirko_charge_2003} calculations, our model yields unphysical astronomical results with cloud columns much greater than $A_\mathrm{V}({\mathrm{tot}})$ of 3 and even larger CRIR than presently inferred. If the \cite{spirko_charge_2003} calculations are correct, we would need to reevaluate our code and update our conclusions. Further theoretical and experimental charge exchange studies are highly desirable to help resolve the issue. Whatever the resolution, our improved OH$^+$ DR rate coefficient will be critical for reliably determining the CRIR from observations of OH$^+$ and H$_2$O$^+$.

Our new estimate of the CRIR derived from OH$^+$ and H$_2$O$^+$ observations in diffuse atomic clouds is now $2.5$ times larger than that inferred from an analysis of H$_3^+$ observations in diffuse molecular clouds using the updated model, whereas the previous study of \cite{neufeld_cosmic_2017} concluded that the CRIR derived from OH$^+$ and H$_2$O$^+$ was in good agreement with that derived from H$_3^+$. This difference suggests that cosmic rays are attenuated as they penetrate from the atomic into the molecular regions of diffuse clouds, where H$_3^+$ is most abundant. This, in turn, could constrain the low energy spectrum of cosmic rays, since the effects of shielding are most pronounced at low energies \citep[e.g.,][]{silsbee_diffusive_2019}. Such shielding effects have been hypothesized from the fact that the CRIR inferred for dense molecular clouds is lower by an order of magnitude or more than that determined in diffuse molecular clouds from observations of H$_3^+$ \citep[e.g.,][]{indriolo_investigating_2012}. To confirm our finding, future determinations of the DR rate coefficients for rotationally cold H$_3^+$ and H$_2$O$^+$ will be needed to refine our estimates of the CRIR determined in these different environments.

\section{Summary} \label{sec:summ}

Accurate CRIR values inferred from OH$^+$ observations for the outer layers of diffuse clouds require reliable DR data for OH$^+$ in its ground electronic, vibrational, and rotational level. Here, we have reported the first DR measurement for OH$^+$ in its lowest energy levels. Using these data, we have generated a kinetic temperature rate coefficient suitable for diffuse cloud chemical models (i.e., for OH$^+$ in its ground rotational level). Our results are valid for gas kinetic temperatures of $10$--$20,\!000$~K. Fit parameters for the analytical representations of the kinetic temperature rate coefficient are given in Tables~\ref{tab:RateOldA} and \ref{tab:rateKIDA}. The rate coefficient reported here is suitable for modeling a range of molecular environments in the ISM, such as diffuse cloud and photodissociation regions, and is especially important for the on-going analysis of OH$^+$ observations along diffuse sightlines, e.g., from the SOFIA HyGal survey \citep{jacob_hygal_2022}.

We have explored the astrophysical implications of our experimental results in two parts. First, by adopting the single-zone diffuse cloud model of \cite{bacalla_edibles_2019} and calculating the multiplicative scaling factor for their previously inferred values of the CRIR due to our present DR rate coefficient. Second, by incorporating our experimentally generated DR rate coefficient into the model of \cite{neufeld_cosmic_2017}, along with additional updates to their model. Our combined experimental and modeling results point to a significant increase in the CRIR estimated from observations of OH$^+$, which has important implications for the shielding of cosmic rays in the ISM. Within diffuse clouds, we find that the CRIR determined from OH$^+$ and H$_2$O$^+$ observations is larger by a factor of $2.5$ than that inferred from observations of H$_3^+$, suggesting that significant CR shielding is possible in the transition from diffuse atomic to diffuse molecular clouds.

\section{Acknowledgements}

Financial support by the Max Planck Society is acknowledged. A. K., D. A. N., D. P., D. W. S., and M. G. W. were supported in part by the NASA Astrophysics Research and Analysis program under 80NSSC19K0969. L. W. I. was supported by the Deutsche Forschungsgemeinschaft (DFG, grant no. 431145392).

\newpage

\appendix 

\section{Merged-beams Experiment} \label{app:mbexp}

OH$^+$ was generated in a standard Penning ion source using an $\mathrm{H}_2/\mathrm{O}_2$ gas mixture. A current of up to a few hundred nA was extracted, accelerated to an energy of $\approx 280$~keV, and mass-to-charge selected using a series of dipole magnets, prior to injection into CSR. The number of injected ions was varied between $1\times10^5$ and $2\times10^6$, in order to remain within the linear counting regime of our neutral particle detector. The injected ions were stored on a closed orbit in CSR and merged with a magnetically confined electron beam in one of the straight sections of the ring. The storage ring lattice was set up in a configuration where the momentum dispersion, i.e., the coupling between ion orbit and momentum, was near zero in the electron--ion overlap section, which improves the electron cooling capabilities of CSR.

The electron beam can be used to cool the stored ions through phase-space cooling (explained below) and through internal-state-changing collisions (see Appendix~\ref{app:rotmodel}), as well as a target for collision studies. Here, we generated a beam current of $15$ or $3.75$~{\textmu}A, in both cases with a circular density profile that was nearly uniform. We will refer to these cases as high or low electron current DR measurements, respectively. The electrons were initially electrostatically accelerated to an energy of $\approx 30$~eV and magnetically guided into CSR, following a step-wise decreasing magnetic field. This produced an adiabatic expansion of the beam, increasing its size and simultaneously reducing the energy spread perpendicular to the magnetic field lines. The effective interaction region of the merged-beams setup is defined by a set of biased drift tubes that control the laboratory-frame energy of the electron beam. Within this region, we applied the lowest guiding field ($10$~mT). This resulted in an expansion factor of $30$ and an effective beam diameter of $12.4\pm0.6$~mm for the high current and $10.2\pm0.9$~mm for the low current measurements. For the perpendicular electron-beam temperature, we take the $k_\mathrm{B}T_\perp  = 2.0^{+1.0}_{-0.5}$~meV estimate for an expansion factor of $20$ of \cite{paul_dr_2022} and extend its lower limit (i.e., $2.0 \pm 1.0$~meV) to account for the factor of $1.5$ times larger magnetic expansion in the present experiment.

Elastic collisions between the electrons and ions enabled us to reduce the size and energy spread of the stored ion beam by a process known as phase-space or electron cooling \citep{poth_cooling_1990}. We achieved electron cooling by matching the average laboratory-frame electron-beam velocity to that of the stored ions. Observing the revolution frequency and momentum spread of the stored ions by means of a Schottky pickup \citep{vonhahn_csr_2016}, we have verified the condition of matched velocities and derived the corresponding laboratory-frame electron-beam energy $E_0 = 9.029\pm0.013$~eV, also called the cooling energy. During collision experiments, we applied electron cooling immediately after ion injection. For the high current measurements, within $10$~s, this resulted in the reduction of the injected ion-beam horizontal and vertical FHWM to $<4.7\pm0.6$~mm and $<3.5\pm0.4$~mm, respectively, in the interaction region. For the low current measurements, these ion-beam FWHM limits were reached within $30$~s.

Using the electron beam as a collision target, we probed the DR energy dependence by detuning the electron-beam energy $E_\mathrm{e}$ from matched velocities. This was done by varying the voltage on the drift tubes. For monoenergetic beams, the resulting collision energy is given by Equation~(\ref{eq:det}). We measured for predefined sets of $E_\mathrm{d}$ values and recorded the corresponding DR product rates. The neutral DR products were collected using a position-sensitive particle-counting detector downstream from the interaction region. We operated the detector in fragment-imaging mode, which enabled us to determine the kinetic energy released in the observed DR process and to image the ion-beam profile. Data for a given $E_\mathrm{d}$ were collected for $25$~ms, followed by electron cooling for $100$~ms, a reference energy at $E_\mathrm{d} = 10$~eV for $25$~ms (used for consistency checks), and electrons off for $25$~ms. We also implemented an $\sim 5$~ms waiting time between each of these steps so that the system could stabilize. The experimental merged-beams rate coefficient $\alpha^\mathrm{mb}$ is proportional to the measured DR rate, as is explained next.

\section{Merged-beams Rate Coefficient} \label{app:absscale}

The measured absolute merged-beams rate coefficient is given by
\begin{equation}
    \alpha^\mathrm{mb}(E_\mathrm{d}) = \frac{R_\mathrm{e}(E_\mathrm{d})}{\eta_\mathrm{d}(E_\mathrm{d}) \xi N_\mathrm{i} n_\mathrm{e}(E_\mathrm{d}) \hat{l}_0/C_0}.
\end{equation}
$R_\mathrm{e}(E_\mathrm{d})$ is the electron-induced count rate due to DR and, above $5$~eV, partly also due to DE. The quantity $\eta_\mathrm{d}(E_\mathrm{d})$ is our detection efficiency for DR events. The parameter $\xi$ is the fraction of ions enclosed by the electron beam in the effective interaction region. Here, $\xi > 0.997$ was determined from the profiles of the ion and electron beam for all the measurements. $N_\mathrm{i}$ is the number of stored ions. The electron density is $n_\mathrm{e}(E_\mathrm{d})$. The effective electron--ion overlap length is $\hat{l}_0 = 0.77\pm0.01$~m and its definition is described in the supplemental material of \cite{kalosi_inelastic_2022}. $C_0 = 35.12\pm0.05$~m is the circumference of the ion-beam orbit.

$R_\mathrm{e}(E_\mathrm{d})$ is given by the difference between the count rates for the measurement and electrons-off steps. The electrons-off step measures the background count rate due to residual gas induced collisions and the intrinsic dark rate of the neutral detector. Reaction studies have shown that turning the electron beam on and off does not cause fluctuations of the background rate (K\'alosi et al. 2023, in preparation). The dark rate of the neutral detector was measured before ion injection.

We determined $\eta_\mathrm{d}(E_\mathrm{d})$ using the approach of \cite{paul_dr_2022} to compare single-to-double particle hits on the detector. This parameter is dependent on the kinetic energy released in the DR process and the branching ratios between open dissociation channels. Both of these can vary as a function of $E_\mathrm{d}$.
At matched velocities, we obtained $\eta_\mathrm{d}(0~\mathrm{eV}) = 0.690$. We probed the energy dependence $\eta_\mathrm{d}(E_\mathrm{d})$ at a few dedicated energies between $1$ and $10$~meV. The values ranged between $0.675$ and $0.690$. In the $10$ to few hundred meV range, in order to have sufficient counting statistics, we subdivided and averaged our data over 4 ranges and found values within the above limits. Finally, we characterized $\eta_\mathrm{d}$ around $1.6$~eV, below which the $\mathrm{O}(^1\mathrm{D}) + \mathrm{H}(n=1)$ DR channel dominates and above which the $\mathrm{O}(^3\mathrm{P}) + \mathrm{H}(n=2)$ channel opens \citep{stromholm_imaging_1997}. Probing at selected energies from $1.5$~eV to $1.9$~eV, $\eta_\mathrm{d}(E_\mathrm{d})$ increased monotonically from $0.640$ to $0.720$. For our work here, we used a single effective $\eta_\mathrm{d} = 0.68\pm0.01$ for the entire energy range studied. This introduces a negligible systematic error for $E_\mathrm{d} \gtrsim 1$~eV compared to the statistical counting uncertainties at those energies.

The merged-beams rate coefficient was initially analyzed on a relative scale using a proxy for the storage-time dependent ion number $N_\mathrm{i}(t)$. In specific, we used the residual gas induced count rate $R_\mathrm{g}(t)$, determined from the electrons-off rate minus the dark rate. $R_\mathrm{g}$ results from collisions of the stored ions with residual gas, generating neutral products through either fragmentation or charge exchange. These measurements were followed by a calibration of $R_\mathrm{g}(t)$ to $N_\mathrm{i}(t)$, expressed as a proportionality factor $S_\mathrm{b} = R_\mathrm{g}(t)/N_\mathrm{i}(t)$. $N_\mathrm{i}$ was measured using beam bunching combined with a capacitive current pickup \citep[PU-C; ][]{vonhahn_csr_2016}. The induced voltage conversion to current for the PU-C was calibrated by \cite{paul_dr_2022} with a $10\%$ systematic uncertainty. $S_\mathrm{b}$ is specific to the present OH$^+$ campaign as it is proportional to residual gas density and collisional rate coefficient for OH$^+$ on the residual gas. We applied ion-beam bunching for durations of $0.5$~s at various storage times and for ion numbers in the range of $2 \times 10^5$ to $1 \times 10^6$ to verify the linear behaviour of $S_\mathrm{b}$. The combined uncertainty due to counting statistics and the PU-C voltage-signal quality contributes an additional $10\%$ uncertainty to the determination of $S_\mathrm{b}$. In addition, we also monitored the relative DR rate coefficient over the energy range $E_\mathrm{d} < 30$~meV. These DR measurements were performed over the course of the OH$^+$ DR campaign. From a comparison of the corresponding relative rate coefficients, we were able to verify that the residual gas density remained constant during the campaign, to within the statistical counting uncertainties of the measurements.

The energy dependent $n_\mathrm{e}(E_\mathrm{d})$ was determined from the measured electron-beam current and radius, accounting for the laboratory-frame electron energy versus $E_\mathrm{d}$. The high current measurement value at matched velocities was $n_\mathrm{e}(0~\mathrm{eV}) = (4.4\pm0.4)\times10^{5}$ cm$^{-3}$. For the low current measurements, $n_\mathrm{e}(0~\mathrm{eV}) = (1.6\pm0.3)\times10^{5}$ cm$^{-3}$.

The total systematic uncertainty of the absolute scaling for our merged-beams DR rate coefficient $\alpha^\mathrm{mb}$ was determined by three dominant contributions. These are the uncertainties of the PU-C voltage-to-current calibration factor, the simultaneous measurement of residual gas induced count rate and bunched-beam induced voltage, and the electron-beam density. All contributing uncertainties were treated as random sign errors and added in quadrature, resulting in a total systematic uncertainty of $17\%$ for the high current measurements and $25\%$ for the low current measurements.

\section{Rotational Level Population Evolution Model} \label{app:rotmodel}

We have constructed a detailed collisional-radiative model in order to predict the rotational populations for the stored OH$^+$ ions. A detailed description of the underlying rate equations is given in the supplemental material of \cite{kalosi_inelastic_2022}, who constructed a similar model for CH$^+$.

The OH$^+$ from our ion source are electronically, vibrationally, and rotationally excited. The low-lying excited electronic states of OH$^+$ decay on timescales $< 1$~ms, except for the first excited a~$^{1}\Delta$ state, which has a calculated radiative lifetime to the X~$^{3}\Sigma^{-}$ ground electronic state on the order of $\sim 30$~ms \citep{stromholm_imaging_1997}. \cite{werner_dipole_1983} calculated vibrational transition probabilities within the X~$^{3}\Sigma^{-}$ ground electronic state, finding radiative lifetimes $< 5$~ms. Given the timescale of our experiments on the order of several tens to hundreds of seconds, we can safely assume that the ions for our measurements were in their ground electronic and vibrational states.

The rotational structure of X~$^{3}\Sigma^{-}$ OH$^+$ is best represented by the quantum number $N \geq 0$, analogous to the rotational ladder of a $^{1}\Sigma$ molecule. Each level with $N>0$ is split into three $J$ sublevels by spin-spin and spin-rotation interactions. Each $J$ sublevel lies $< 3$~cm$^{-1}$ (4.3~K) away from the $N$ level rotational energy \citep{hechtfischer_nearthreshold_2019}. The three $J$ sublevels for a given $N$ have practically equal radiative lifetimes. This substructure was omitted in the final model as radiative cooling in the CSR radiation field, when including the sublevels, leads to equilibrium $J$ populations similar to an $\approx 17$~K Boltzmann distribution. This temperature is sufficiently high compared to the level splittings so that the sublevels for each $N$ are statistically populated with weights $2J+1$. This enabled us to neglect the sublevels in our final model for the time evolution of the $N$ level populations.

The approximate rotational energies of OH$^+$ X~$^{3}\Sigma^{-}(v=0)$ were calculated using the $B_0$ and $D_0$ spectroscopic constants of \cite{hodges_fourier_2017} as
\begin{equation} \label{eq:rotlevel}
    E_{N} = B_0 [N(N+1)] - D_0 [N(N+1)]^2.
\end{equation}
Levels up to $N=19$ were included in the model. We used the theoretical $2.26$~D dipole moment of \cite{werner_dipole_1983} to calculate Einstein coefficients for spontaneous emission for rotational transitions using Equation~(S1) from \cite{kalosi_inelastic_2022}, with $J$ replaced by $N$.

Our radiative cooling model also accounts for the ambient radiation field in CSR, following the approach of \cite{meyer_radiative_2017}. Their model consists of two components: one for the thermal radiation of the cryogenic chambers and the second for room-temperature leaks from various openings in CSR. Here, we estimated the cryogenic component to be $T_\mathrm{low} \approx 6$~K, based on the measured chamber temperatures. The room-temperature component is fixed to $300$~K. Previous work has found its fraction to be $\varepsilon = (1.0 \pm 0.3)\times10^{-2}$ \citep{kalosi_inelastic_2022}. Of these two components, the room-temperature fraction dominates the accuracy of our rotational population predictions when approaching equilibrium with the CSR radiation field.

Rotational-level-changing collisions between electrons and ions have been experimentally demonstrated for CH$^+$ to affect the rotational level populations in our experimental setup \citep{kalosi_inelastic_2022}. To examine the role of these collisions for OH$^+$, we adopted the theoretical electron-impact rotational excitation cross section of \cite{hamilton_electronimpact_2018} for $\Delta N = 1$ and $2$ transitions of OH$^+$. The corresponding de-excitation cross sections were obtained by applying the principle of detailed balance. We then calculated rotational-level-changing merged-beams rate coefficients for all $E_\mathrm{d}$ corresponding to those used for our DR rate coefficient measurements. For a given transition, the importance of rotational-level-changing collisions can be shown by the ratio of the Einstein $A_{N^{\prime} \rightarrow N^{\prime\prime}}$ rate to the $N$-level changing merged-beams rate coefficient $\alpha_{N^{\prime} \rightarrow N^{\prime\prime}}$, which give the critical electron density
\begin{equation}
    n_\mathrm{c} = A_{N^{\prime} \rightarrow N^{\prime\prime}}/\alpha_{N^{\prime} \rightarrow N^{\prime\prime}}.
\end{equation}
Our DR measurements here were carried out for ions primarily in the $N=0$ and $1$ levels. For the $N = 1 \rightarrow N = 0$ transition, we calculated $A_{1 \rightarrow 0} = 1.89\times10^{-2}$~s$^{-1}$ and a critical electron density of $(4.5 \pm 0.8) \times 10^{3}$~cm$^{-3}$ at matched velocities. This value is comparable to the typical ring-averaged electron density of $(9.6 \pm 0.9) \times 10^{3}$~cm$^{-3}$ for the high current measurements, meaning that both collisions and radiative interactions are important in our experiment for the $N=0$ and $1$ levels.

\subsection{Results for High Electron Current DR Measurements} \label{app:fixpop}

\begin{figure}[ht!]
\plotone{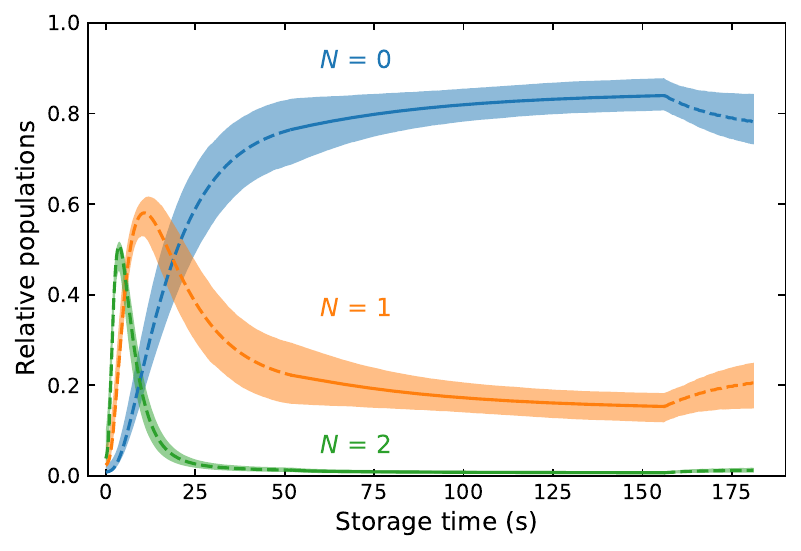}
\caption{Collisional-radiative model of OH$^+$ rotational level population evolution as a function of ion storage time for the high electron current DR measurements. The dashed and solid lines are for the mean model with the parameters $k_\mathrm{B}T_\perp = 2$~meV, $T_\mathrm{low} = 6$~K, $\varepsilon = 1.0\times10^{-2}$, and $3000$~K initial rotational temperature. The accompanying shaded areas indicate the uncertainty of the predictions when varying these parameters, the dipole moment, the electron-impact rotational excitation cross sections, and the initial rotational temperature, all within their estimated uncertainties. The dashed or solid lines differentiate between periods with or without electrons, respectively. \label{fig:rotmod}}
\end{figure}

The high electron current measurements were designed to collect statistically significant DR signal compared to background at all collision energies of interest, within the two weeks of the measurement campaign. The complete population model, including both radiative cooling in the CSR radiation field and rotational-level-changing collisions, was used to optimize the internal-state preparation scheme for the high current DR measurements, for which the measured data are shown in Figure~\ref{fig:mbrc}. The scheme consisted of three ion storage phases. First, following injection, we applied electron cooling at matched velocities for $52$~s. During this phase, rotational-level-changing collisions accelerated the rotational cooling of the stored ions compared to only radiative cooling. Next, we turned off the electron beam for $104$~s to equilibrate the rotational level populations with the CSR radiation field. During this phase, the FWHM of the ion beam profile slowly increased, due to ion intrabeam heating processes. Finally, we measured for predefined sets of $E_\mathrm{d}$ for $25$~s. The $\alpha^\mathrm{mb}$ data shown in Figure~\ref{fig:mbrc} have been evaluated after excluding the first $3$~s of the measurement phase where the ion beam FWHM was decreasing, due to electron cooling.

The modeled relative populations are shown in Figure~\ref{fig:rotmod} for the levels of interest ($N \leq 2$) as a function of storage time from injection. The initial rotational level populations are represented by a Boltzmann distribution within the ground vibrational state. The final population results for levels $N \leq 2$ are insensitive to the initial populations for an assumed initial rotational temperature $> 300$~K. In the initial cooling phase, the time needed for the $N=0$ population to approach its end value is reduced by a factor of three compared to only radiative cooling. After turning off the electron current, the $N=0$ population continues to grow as the system approaches equilibrium with the CSR radiation field. In the measurement phase, the populations are determined by the combined effects of collisions and radiation. The additional cycling between the selected $E_\mathrm{d}$ and accompanying electron cooling, reference, and electron-off steps results in a slight increase of the effective rotational temperature.

 The model uncertainties were determined primarily by varying the experimental parameters $T_\perp$, $T_\mathrm{low}$ and $\varepsilon$. Two additional parameters can affect the model predictions, the dipole moment and the electron-impact rotational excitation cross sections for OH$^+$. The magnitudes of these two parameters primarily affect the time scale of the rotational population evolution. The dipole moment is expected to be accurate within $7 \%$, based on a comparison of the same level of theory and a measurement for OH$^-$ \citep{meyer_radiative_2017}. The theoretical model used by \cite{hamilton_electronimpact_2018} to calculate the cross sections for rotational-state changing collisions has been experimentally benchmarked for CH$^+$ by \cite{kalosi_inelastic_2022} and good agreement was found with theory to better than $\sim 40\%$. For the present high electron current results, when the populations are near equilibrium, the uncertainties in the dipole moment and the collisional cross sections do not contribute significantly to the uncertainty budget.

\begin{figure}[ht!]
\plotone{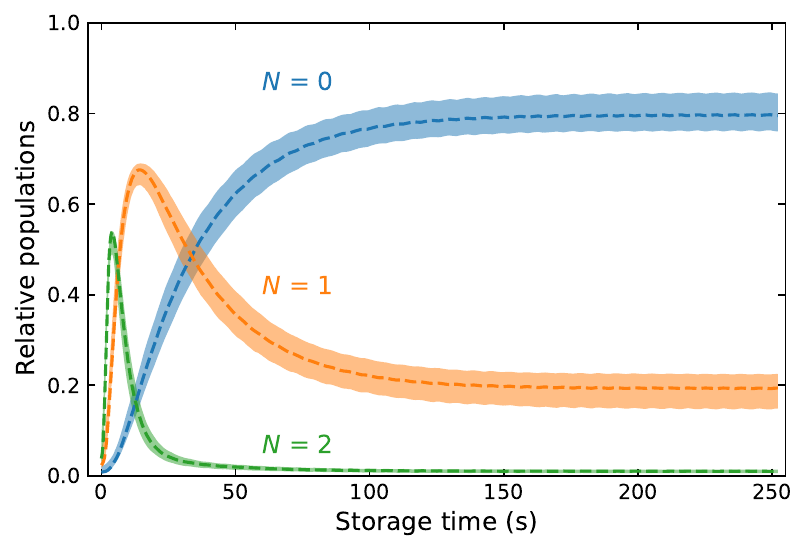}
\caption{Same as Figure~\ref{fig:rotmod}, but for the low electron current DR measurements. \label{fig:rottime}}
\end{figure}

The model uncertainty determination uses a Monte Carlo approach. The dipole moment, electron impact rotational excitation cross sections, and $\varepsilon$ are drawn from normal distributions. For the cross sections, we assumed a systematic uncertainty of $40\%$ at all energies and for all levels. The initial rotational temperature, $T_\mathrm{low}$, and $T_\perp$ are drawn from uniform distributions to include the physical limits on these parameters. For the initial rotational temperature, an upper limit of $5000$~K was chosen. For $T_\mathrm{low}$, the lower and upper limits were set to $4$ and $8$~K, respectively. At each storage time value, we evaluated the 16th and 84th percentiles of the population distributions, which is equivalent to $\pm$~one-sigma for a normal distribution. The shaded areas in Figure~\ref{fig:rotmod} are given for these percentiles.

Using our model, we estimate the average relative populations of rotational levels in the measurement time window to be $0.80\pm0.05$ for $N=0$, $0.19\pm0.04$ for $N=1$, and $\approx 0.01$ for $N=2$. The uncertainties for the two lowest $N$ levels are anti-correlated and the sum of their populations totals to $\approx 0.99$.

\subsection{Results for Low Electron Current DR Measurements} \label{app:timepop}

The low electron current measurements were designed to reduce the influence of rotational-level-changing collisions. The corresponding ring-averaged electron density was $(3.5 \pm 0.7) \times 10^{3}$~cm$^{-3}$ at matched velocities. These low current measurements were performed by cycling between the selected $E_\mathrm{d}$ and the accompanying electron cooling, reference, and electron-off steps, starting immediately after injection. The relative population results of the model are shown in Figure~\ref{fig:rottime} for $N \leq 2$. The initial rotational level populations are represented by a Boltzmann distribution within the ground vibrational state with an assumed initial rotational temperature of $> 300$~K.

The model uncertainties were determined using the Monte Carlo approach described above. Compared to the results shown in Figure~\ref{fig:rotmod}, the uncertainty of the predictions is smaller at storage times where the $N=0$ population is increasing, which enables us to more reliably extract the level-specific $\alpha^\mathrm{mb}_N$ for $N=0$ than when using a higher electron current.

\section{$N$-Level Specific DR Measurements} \label{app:vststorm}

The determination of level-specific $\alpha^\mathrm{mb}_{N}$ values is enabled by the time evolution of the rotational level populations for stored ions. Here, we have determined $\alpha^\mathrm{mb}_{N}$ using the data from our low electron current DR measurements.

\begin{figure}[ht!]
\plotone{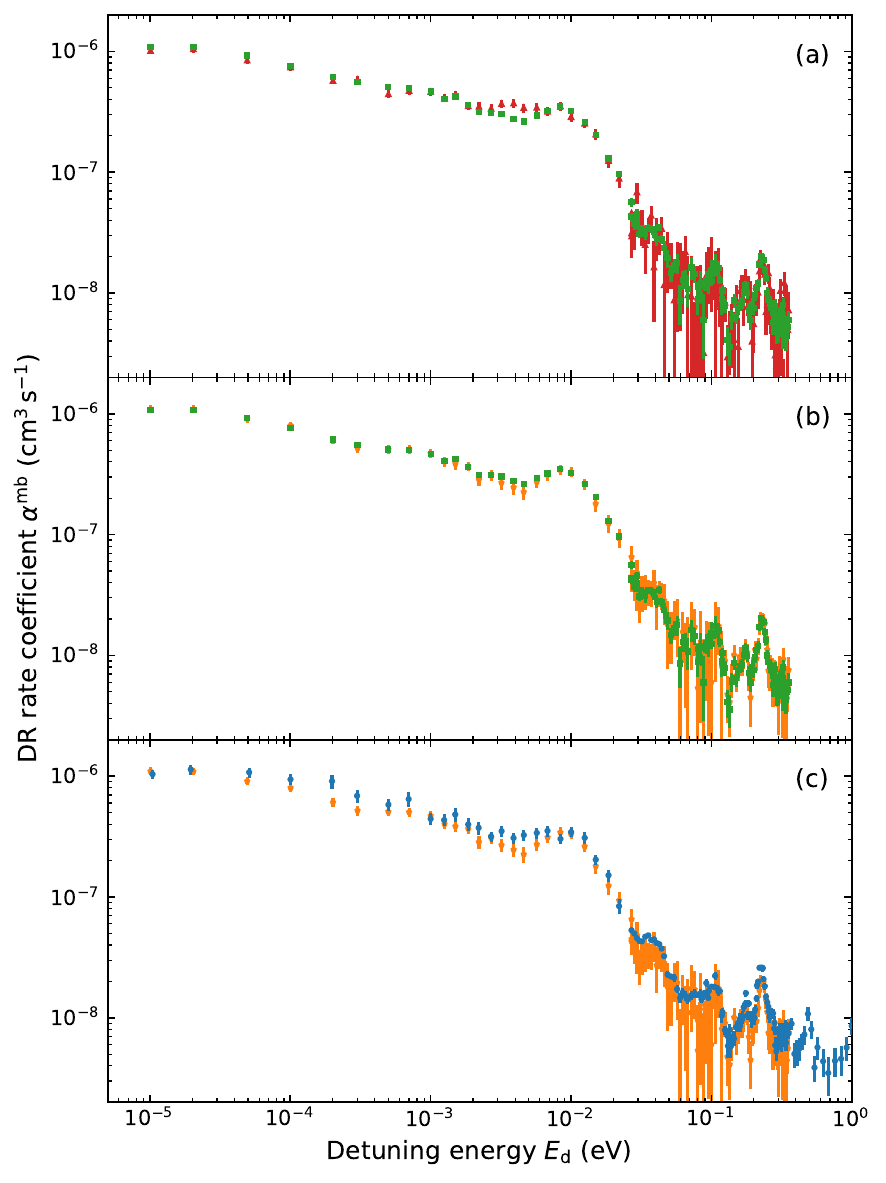}
\caption{OH$^+$ DR rate coefficient $\alpha^\mathrm{mb}(E_\mathrm{d})$ for low electron current measurements. (a) Comparison of $\alpha^\mathrm{mb}$ from two storage-time windows, from $\approx25$ to $\approx33$~s (red triangles) and from $\approx145$ to $\approx250$~s (green squares). (b) Comparison of $\alpha^\mathrm{mb}$ from the storage-time window with equilibrated populations (from $\approx145$ to $\approx250$~s, green squares) and the fitted $\alpha^\mathrm{mb}_{0}$ (orange pentagons). (c) Comparison of $\alpha^\mathrm{mb}$ from the high electron current measurements (blue circles) with the fitted $\alpha^\mathrm{mb}_{0}$ (orange pentagons). All error bars represent one-sigma statistical uncertainties. \label{fig:tstorm}}
\end{figure}

The measured $\alpha^\mathrm{mb}(t)$ for a given storage-time window $t$ is the sum
\begin{equation} \label{eq:ratetstorm}
    \alpha^\mathrm{mb}(t) = \sum_{N} \bar p_{N}(t)\, \alpha^\mathrm{mb}_{N},
\end{equation}
where the rotational-level-specific rate coefficients $\alpha^\mathrm{mb}_{N}$ are weighted by the average populations $\bar p_{N}(t)$ in the time window. The results shown in Figure~\ref{fig:rottime} were used to calculate $\bar p_{N}(t)$. By including more storage-time windows than the number of contributing rotational levels, we can perform a least squares fit to Equation~(\ref{eq:ratetstorm}) and extract $\alpha^\mathrm{mb}_{N}$.

The earliest storage-time window included in the analysis spans from $\approx25$ to $\approx33$~s, followed by several windows of the same length. Earlier storage times where the ion beam FWHM was decreasing due to electron cooling were excluded. Based on our collisional-radiative model results shown in Figure~\ref{fig:rottime}, rotational equilibrium was reached after $\approx 150$~s of ion storage. The final storage-time window was extended to cover the times from $\approx145$ to $\approx250$~s. Only levels with $>1\%$ relative population ($N \leq 2$) were included in the fit. In Figure~\ref{fig:tstorm}(a), we compare the measured $\alpha^\mathrm{mb}$ for two selected storage-time windows in the experiment: from $\approx25$ to $\approx33$~s and $\approx145$ to $\approx250$~s.

In order to include the uncertainty of the collisional-radiative model in the level-specific results, we used our Monte Carlo model to generate a set of $\bar p_{N}(t)$, as described above. A fit by Equation~(\ref{eq:ratetstorm}) was performed for each $\bar p_{N}(t)$ within the set. In the selected time windows, the $N = 2$ population was small but non negligible. This led to an uncertainty for $\alpha^\mathrm{mb}_{2}$ that was much larger than the measured rate coefficient at any storage time. Due to a strong correlation between the $N = 2$ and $1$ populations, this uncertainty propagated into the $\alpha^\mathrm{mb}_{1}$ result as well. The combination of the large $\alpha^\mathrm{mb}_{2}$ uncertainty and the small differences between the measured $\alpha^\mathrm{mb}$ for the sampled storage time windows, see Figure~\ref{fig:tstorm}(a), led to an uncertainty for the $\alpha^\mathrm{mb}_{1}$ result that was also larger than the measured rate coefficient at all storage times. However, for $\alpha^\mathrm{mb}_{0}$, each fit yielded values $\alpha^\mathrm{mb}_{0,i}$ with uncertainties $\delta^\mathrm{mb}_{0,i}$ smaller than the fitted values. The final $\alpha^\mathrm{mb}_{0}$ shown in Figure~\ref{fig:tstorm} was calculated as the unweighted mean of the individual fits $\alpha^\mathrm{mb}_{0,i}$ and its statistical uncertainty as the unweighted mean of the individual fit uncertainties $\delta^\mathrm{mb}_{0,i}$. The variance of the individual fits $s^\mathrm{mb}_{0}$ was much smaller than any of the individual fit uncertainties $\delta^\mathrm{mb}_{0,i}$.

The final $\alpha^\mathrm{mb}_{0}$ of the fits by Equation~(\ref{eq:ratetstorm}) for the Monte Carlo generated $\bar p_{N}(t)$ is plotted in Figure~\ref{fig:tstorm}(b) and (c). Figure~\ref{fig:tstorm}(b) compares the extracted $\alpha^\mathrm{mb}_{0}$ to the data from the final storage-time window of the low electron current measurements. Figure~\ref{fig:tstorm}(c) compares the extracted $\alpha^\mathrm{mb}_{0}$ to the data from the high current measurements, i.e., from Figure~\ref{fig:mbrc}. The present analysis shows that the measured rate coefficient for our high electron current DR measurements well approximates $\alpha^\mathrm{mb}_{0}$. When converting the measured data into $\alpha^\mathrm{k}$, the observed differences become negligible. Hence, $\alpha^\mathrm{k}(T_\mathrm{k})$ shown in Figure~\ref{fig:tkin} is well suited for chemical kinetics models for ground rotational level OH$^+$.

\section{Kinetic Temperature DR Rate Coefficient} \label{app:ratecoeff}

Here, we provide two different analytic representations for the OH$^+$ DR kinetic temperature rate coefficient $\alpha^\mathrm{k}$ and its one-sigma error band shown in Figure~\ref{fig:tkin}. First, we follow \cite{novotny_drhcl_2013} and fit $\alpha^\mathrm{k}$ with an optimized function, with a small number of fit parameters, that accounts for typical DR features, i.e., broad peaks or dips from resonances. The fit function is given by
\begin{eqnarray} \label{eq:RateFunctionON}
	\alpha^{\mathrm{k}}(T_{\mathrm{k}})[\mathrm{cm}^3\, \mathrm{s}^{-1}]&=&
	A\left(\frac{300}{T_{\mathrm{k}}[\mathrm{K}]}\right)^n\\ 
	&+&T_{\mathrm{k}}[\mathrm{K}]^{-1.5}\sum_{r=1}^{4}c_r\exp\left(-\frac{T_r}{T_{\mathrm{k}}[\mathrm{K}] }\right), \nonumber
\end{eqnarray}
and the parameters are listed in Table~\ref{tab:RateOldA}. The maximum relative deviation of the fit is $0.5\%$ below $6000$~K and  $1.4\%$ from there to $20,000$~K.

\begin{deluxetable}{cccc} \label{tab:RateOldA}
	\tabletypesize{\scriptsize}
	\tablecaption{Fit Parameters for the OH$^{+}$ Kinetic Temperature Rate Coefficient $\alpha^\mathrm{k}$ and Its Lower and Upper Error Band from Figure~\ref{fig:tkin}, Using Equation~(\ref{eq:RateFunctionON}).}
	\tablewidth{0pt}
	\tablehead{
		\colhead{Parameter} & \colhead{Rate coefficient} & \colhead{Lower error limit} & \colhead{Upper error limit}
	}
	\startdata
	$A$       & ${1.10\times 10^{-7}}$       & ${9.76\times 10^{-8}}$    & ${1.32\times 10^{-7}}$        \\ 
	$n$       & ${0.767}$          & ${0.655}$      & ${0.812}$          \\ 
	$c_1$     & ${3.46\times 10^{-4}}$  & ${1.88\times 10^{-4}}$  & ${4.76\times 10^{-4}}$  \\ 
	$c_2$     & ${-9.16\times 10^{-4}}$  & ${-9.18\times 10^{-4}}$  & ${-8.19\times 10^{-4}}$  \\ 
	$c_3$     & ${-1.85\times 10^{-3}}$  & ${-2.31\times 10^{-3}}$  & ${-7.32\times 10^{-4}}$  \\ 
	$c_4$     & ${3.33\times 10^{-2}}$  & ${-6.83\times 10^{-3}}$  & ${4.06\times 10^{-2}}$   \\ 
	$T_1$     & ${129}$         & ${114}$         & ${137}$         \\ 
	$T_2$     & ${1220}$        & ${868}$        & ${1290}$        \\ 
	$T_3$     & ${5900}$        & ${2980}$        & ${6990}$        \\ 
	$T_4$     & ${43400}$       & ${10100}$       & ${37500}$       \\
	\enddata
\end{deluxetable}

Second, we give the Arrhenius--Kooij (AK) representation, as is typically used in astrochemistry, combustion chemistry, and other chemical models and databases. Following the approach of \cite{paul_dr_2022}, we provide a set of piecewise-joined fit functions on several temperature intervals, since we cannot model the experimental results with a single AK fit function. This approach introduces discontinuities in the temperature dependence of the analytical rate coefficient between the temperature intervals, which can be avoided by using the representation by Eq.~(\ref{eq:RateFunctionON}). The temperature-interval fit function is given by
\begin{equation} \label{eq:RateFitKIDA}
	\alpha^{\mathrm{k}}(T_{\mathrm{k}})[\mathrm{cm}^3\, \mathrm{s}^{-1}]=A\left(\frac{T_{\mathrm{k}}[\mathrm{K}]}{300}\right)^{\beta} e^{-\frac{\gamma}{T_{\mathrm{k}}[\mathrm{K}]}},
\end{equation}
and the parameters for each temperature interval are listed in Table~\ref{tab:rateKIDA}. The maximum relative deviation of the fit is $1\%$. Following KIDA conventions, the relative uncertainty is described by the log-normal factor $F = \exp{\left(\Delta \alpha^\mathrm{k}/\alpha^\mathrm{k}\right)} $. This quantity is fitted on the same temperature intervals as for the AK fits by the fit function 
\begin{equation} \label{eq:RateUncertaintyFitKIDA}
	F(T_{\mathrm{k}})=F_{0}\exp\left(g\left(\frac{1}{T_{\mathrm{k}}[\mathrm{K}]}-\frac{1}{300} \right)\right) .
\end{equation}
Due to the asymmetric error bands of $\alpha^\mathrm{k}$, the log-normal factor is calculated as the average of the upper and lower error bands. The continuity of both the $\alpha^\mathrm{k}$ and error fits are guaranteed on the border of each temperature range.

\begin{deluxetable*}{ccccccc} \label{tab:rateKIDA}
	\tabletypesize{\small}
	\tablewidth{\textwidth}
	\tablecaption{Fit Parameters for the OH$^{+}$ Kinetic Temperature Rate Coefficient $\alpha^\mathrm{k}$ and Its Relative Uncertainty From Figure~\ref{fig:tkin}, Using Equations~(\ref{eq:RateFitKIDA}) and (\ref{eq:RateUncertaintyFitKIDA}), Respectively.}
	\tablehead{
		\colhead{Parameter} & \multicolumn{6}{c}{Temperature range (K)} \\  & \colhead{$10$--$30$} & \colhead{$30$--$100$} & \colhead{$100$--$1000$} & \colhead{$1000$--$4000$} & \colhead{$4000$--$11,000$} & \colhead{$11,000$--$20,000$}
	}
	\startdata
    	$A$       & ${1.58\times 10^{-7}}$       & ${1.98\times 10^{-7}}$    & ${1.78\times 10^{-7}}$  & ${1.02\times 10^{-7}}$ & ${1.48\times 10^{-8}}$ & ${1.98\times 10^{-9}}$      \\ 
	$\beta$       & ${-0.595}$       & ${-0.519}$    & ${-1.099}$  & ${-0.845}$  &  ${-0.301}$  &  ${0.142}$   \\
	$\gamma$       & ${-2.30}$       & ${-0.792}$    & ${52.4}$  & ${-198}$ & ${-2285}$ & ${-6885}$      \\ \hline
	$F_0$      & ${1.32}$       & ${1.27}$    & ${1.28}$  & ${1.31}$ & ${1.28}$ & ${1.40}$      \\
	$g$       & ${1.32}$       & ${2.68}$    & ${0.764}$  & ${8.30}$ & ${2.26}$ & ${29.7}$      \\
	\enddata
\end{deluxetable*}

\bibliography{OH+ApJL}
\bibliographystyle{aasjournal}



\end{document}